\documentclass[aps,prl,twocolumn,superscriptaddress,showpacs]{revtex4}
\usepackage{graphicx}
\usepackage{mathrsfs}
\usepackage{bm}
\usepackage{color}
\usepackage{soul}
\begin{document}

\title{Neutrino footprint in Large Scale Structure}
\author{Raul Jimenez}
 \email{raul.jimenez@icc.ub.edu}
\affiliation{ICREA \& ICC, University of Barcelona, Marti i Franques 1, Barcelona 08028, Spain.} 
\affiliation{Radcliffe Institute for Advanced Study, Harvard University, MA 02138, USA}
\author{Carlos Pena-Garay}
\email{penya@ific.uv.es}
\affiliation{Instituto de Fisica Corpuscular, CSIC-UVEG, P.O.  22085, Valencia, 46071, Spain}
\affiliation{Laboratorio Subterr\'aneo de Canfranc, Estaci\'on de Canfranc, 22880, Spain}
\author{Licia Verde}
\email{liciaverde@icc.ub.edu}
\affiliation{ICREA \& ICC, University of Barcelona, Marti i Franques 1, Barcelona 08028, Spain.} 
\affiliation{Radcliffe Institute for Advanced Study, Harvard University, MA 02138, USA}
\affiliation{Institute of Theoretical Astrophysics, University of Oslo, Oslo 0315, Norway.}

\begin{abstract}
  Recent constrains on the sum of neutrino masses inferred by analyzing cosmological data, show 
  that detecting a non-zero neutrino mass 
  is within reach of  forthcoming cosmological surveys, implying a direct determination of the absolute neutrino mass scale. The measurement relies on constraining the shape of the matter power spectrum below the neutrino free streaming scale: massive neutrinos erase power at these scales. Detection of a lack of small-scale power, however, could also be due to a host of other   effects. It is therefore of paramount importance  to validate neutrinos as the 
  source of power suppression at small scales.  We show that, independent on hierarchy, neutrinos 
  always show a footprint on large, linear scales; the exact location and properties  can be related to the measured power suppression (an astrophysical measurement) and 
  atmospheric neutrinos mass splitting (a neutrino oscillation experiment measurement). This feature can not be easily mimicked  by  systematic 
  uncertainties or modifications in the cosmological model.   The measurement of such a feature, up to 1\% relative change in the power spectrum, is a smoking gun for confirming  the  determination of the absolute neutrino mass scale from cosmological observations. 
  It also demonstrates the synergy of astrophysics and  particle physics experiments.
  
 \end{abstract}
\maketitle


In the past few years, there has been an amazing progress in cosmology.  An accurate cosmic microwave background (CMB) spectrum, 
both in temperature and polarization has been measured by Planck \cite{Ade:2015xua} and WMAP \cite{Hinshaw:2012aka}. The  expansion history of the Universe has been mapped in several ways: with  measurements of  the Baryon 
Acoustic Oscillation (BAO) scale by  the Baryon Oscillation Spectroscopic Survey (BOSS) of the Sloan 
Digital Sky Survey (SDSS) \cite{Anderson:2013zyy} and others \cite{Blake:2011en, Beutler:2011hx}; by the luminosity distance relation as given by Type 1A supernova data  e.g., \cite{Betoule:2014frx}; via the direct measurement of the Hubble parameter with  cosmic chronometers \cite{Moresco:2012jh, Moresco:2016mzx}. 

Finally, large scale structure (LSS) has been probed 
by a variety of surveys 
(galaxies e.g.,  \cite{WZ1,Riemer-Sorensen:2013jsa, Reid:2009xm, Alam:2015mbd}, weak lensing e.g., \cite{DLS, CFHTLens1, Kilbinger:2012qz, Kitching:2014dtq, DES}, Ly$\alpha$ \cite{Palanque-Delabrouille:2015pga})  with increased sensitivity to the scale and redshift dependences of the matter power spectrum, thanks also to redshift  space distortion measurements e.g., \cite{Samushia:2013yga, Beutler:2013yhm}. 

All this wealth of cosmological data show a consistent $\Lambda$CDM model with improved precision on parameters and better control of  
systematics. If included as a parameter in the model, total neutrino mass bounds have significantly improved in a variety of analysis, yielding an upper bound slightly 
higher than 100 meV \cite{Palanque-Delabrouille:2015pga, Cuesta:2015iho}. 
Massive neutrinos free stream out of potential wells, erasing fluctuations and thus suppressing power on small scales  e.g., \cite{Hu:1997mj, Lesgourgues:2006nd}; the measured small scale power is consistent with the standard (massless neutrino) $\Lambda$CDM model and inconsistent with large neutrino masses. 
These bounds are very close to the  limit that separates inverted and normal ordering and is within a 
factor of two of the lower limit of the sum of neutrino masses set by oscillations e.g., \cite{Gonzalez-Garcia:2015qrr}. Near future progress in this measurement has an important impact on a crucial question 
in neutrino physics \cite{paper1}: is neutrino its own anti-particle?  The almost century-old question can be resolved if a neutrinoless double beta decay is observed \cite{Furry:1939qr,Majorana:1937vz}. If light 
neutrinos are the main source of this decay, the measurement of the neutrino mass, provides an estimate of the half-live and therefore the size and level of background of experiments required  to prove it. 

Relevant cosmological constraints on neutrino masses (i.e., coming from measurements of the small scales power suppression)  have to be matched with careful verification 
of systematic effects or alternative explanations of the data. Building on the work of \cite{paper1}, we show that cosmological data contain {\em independent} information 
on the neutrino masses that can be hardly mimicked by other effects. Driven by the  current strong limits in the total mass, we demonstrate that there  is sensitivity to measure the expected large mass splitting of neutrinos if the total mass is measured from forthcoming cosmological  data.  The agreement of the mass splitting inferred from 
cosmology and the large neutrino mass splitting measured by oscillation experiments should prove as convincing evidence of the discovery of 
relic neutrinos in cosmological surveys. 

Massive neutrinos affect  cosmological observations in a variety of different ways. For example, the combination of CMB and BAO data constrain the total neutrino mass $\Sigma<0.23$ eV at the 95\% confidence level \cite{Ade:2015xua}. Neutrinos with mass $\lesssim 1$eV become non-relativistic 
after the epoch of recombination probed by the CMB, thus massive neutrinos alter matter-radiation equality for a fixed $\Omega_m h^2$. After neutrinos become
non-relativistic, their free streaming damps the small-scale  power and modifies the shape of the matter power spectrum below the free-streaming length. 
Combining large-scale structure and CMB data, at present the sum of masses is constrained to be $\Sigma \lesssim 0.13$ eV \cite{Cuesta:2015iho, Palanque-Delabrouille:2015pga}.
Forthcoming large-scale structure data promise to determine the small-scale ($0.1 \lesssim k\lesssim 1$ h/Mpc) matter power spectrum exquisitely well 
and to yield errors on $\Sigma$ well below $0.1$ eV (e.g., \cite{Carbone:2010ik, Audren:2012vy}). Here, we assume the standard $\Lambda$CDM model (plus massive neutrinos) and explore the changes in the
matter power spectra due to the neutrino properties (mass and hierarchy).

Neutrino oscillation data have measured the neutrino squared mass differences, which are hierarchical. Given the smallness of neutrino masses and 
the hierarchy in mass splittings, we can characterize the impact of neutrino masses  on cosmological observables and in particular on 
the matter power spectrum by two parameters: the total mass $\Sigma$ and
the ratio of the largest mass splitting to the total mass, $\Delta$; 
while one can safely neglect the impact of the solar mass splitting in
cosmology. In this excellent approximation, two masses characterize the neutrino mass
spectrum, the lightest one, $m$, and the heaviest one, $M$. 
\begin{center}
\begin{figure}[]
\hspace*{-0.5cm}
\includegraphics[width=9cm]{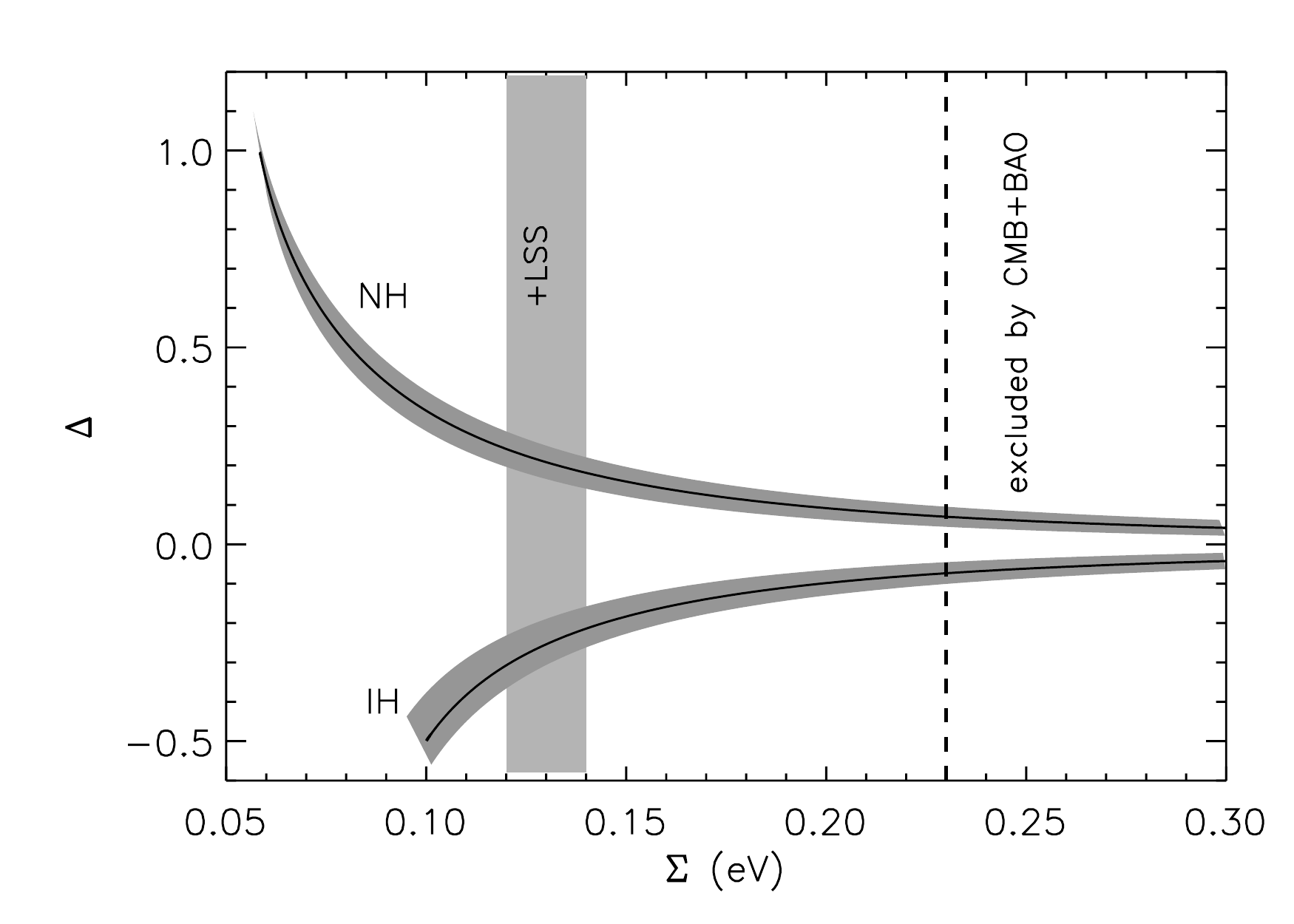}
\caption{Constraints from neutrino oscillations  
(shaded regions) and from cosmology in the $\Sigma$-$\Delta$ plane. In this parameterization the sign of $\Delta$ specifies the hierarchy.} 
\label{fig:0}
\end{figure}
\end{center}
We define the relation between the neutrino masses $m$ and $M$ and the parameters $\Sigma$ and $\Delta$ as 
\begin{eqnarray}
{\rm NH:} \, \, \,\,\,\,\,\, & \Sigma =  2m + M \,\,\,\,\,\, & \Delta=(M-m)/\Sigma \\
{\rm IH:} \, \, \,\,\,\,\,\, & \Sigma =  m + 2M \,\,\,\,\,\, & \Delta=(m-M)/\Sigma \,.
\end{eqnarray} 

 In Fig~\ref{fig:0} we show the current constraints on neutrino mass
 properties in the $\Sigma$-$\Delta$ plane.
 We use the $\Delta$ parameterization for the
 following reasons: 
 $\Delta$ changes continuously through normal, degenerate and inverted
 hierarchies; 
 $\Delta$ is positive for NH and negative for IH;   cosmological data are sensitive to $\Delta$ in an easily understood
 way through the largest mass splitting (i.e., the absolute value of $\Delta$), 
 while the direction of the splitting (the sign of $\Delta$) introduces 
 a sub-dominant correction to the main effect \cite{paper1}. This parameterisation
 is strictly only applicable for $\Sigma > 0$, but oscillations experiments already 
 set $\Sigma>M\gtrsim 0.057$eV.  

It is important to note that not the entire parameter space in the
$\Sigma$-$\Delta$ 
plane (or of any other parameterization of the hierarchy used in the
literature) is allowed  by particle physics
constraints: only the regions
around the normal and inverted 
hierarchies allowed by neutrino oscillation experiments are physical 
(see Fig~\ref{fig:0}).

The effect of neutrino mass on the CMB is related to the
physical density of neutrinos, and therefore the mass difference
between eigenstates can be neglected. However individual neutrino
masses can have an effect on the large-scale shape of the matter power
spectrum. 
In fact, neutrinos of different masses have  different transition
redshifts from 
relativistic to non-relativistic behavior, and their individual masses
and their mass splitting change
the details of the radiation-domination to matter-domination regime. 
As a result, the detailed shape of the matter power spectrum on scales
$k\lesssim 0.01$ $h$/Mpc is affected. Therefore a precise measurement of the matter 
power spectrum shape can give information on both the sum of the
masses and the neutrino mass splitting. 

To gain physical intuition on the effect of neutrino properties 
on cosmological observables, such as the shape of the matter power
spectrum, 
it is useful to adopt the following analytical approximation, 
as described in Ref. \cite{takada}.
The matter power spectrum can be written as:
\begin{equation}
\frac{k^3 P(k;z)}{2 \pi^2} = \Delta_R^2 \frac{2 k^2}{5 H_0^2 \Omega_m^2}  D^2_{\nu} (k,z) T^2(k) \left ( \frac{k}{k_p} \right )^{(n_s-1)},
\end{equation}
where $ \Delta_R^2$ is the primordial amplitude of the fluctuations (evaluated at the pivot scale $k_p$),
$n_s$ is the primordial 
power spectrum spectral slope, $T(k)$ denotes the matter transfer
function and 
$D_{\nu} (k,z)$ is the  scale-dependent linear growth function, which encloses the dependence
of $P(k)$ on non-relativistic neutrino species.

Each of the three neutrinos contributes to the neutrino mass fraction
$f_{\nu,i}$ 
where $i$ runs from $1$ to $3$,
\begin{equation}
f_{\nu,i} = \frac{\Omega_{\nu,i}}{\Omega_m} = f_0 \times  m_{\nu_i} = 0.01 \left ( \frac{m_{\nu_i}}{0.13 {\rm eV}} \right ) \left ( \frac{0.14}{\Omega_m h^2} \right )
\end{equation}
and has a free-streaming scale $k_{{\rm fs},i}$,
\begin{equation}
  k_{{\rm fs},i} = k_0  \sqrt{m_{\nu_i}} = 0.015 \sqrt{\frac{m_{\nu_i}}{0.13 {\rm eV}} \frac{\Omega_m h^2}{0.14} \frac{5}{1+z}}  {\rm Mpc}^{-1}\,.
\end{equation}
Analogously, one can define the corresponding quantities 
for the combined effect of all species, by using $\Sigma$ instead of $m_{\nu_ i}$.
Since we will only distinguish between a light and a heavy eigenstate
we will have e.g., 
$f_{\nu,m}, f_{\nu,\Sigma},  k_{{\rm fs},m},  k_{{\rm fs},\Sigma}$
etc., 
where in the expression for  $f_{\nu,m}$ one should use 
the mass of the eigenstate 
(which is the mass of the individual neutrino, or twice as much 
depending on the hierarchy) while in  $k_{{\rm fs},m}$ one should use
the mass of the 
individual neutrino.

The dependence
of $P(k)$ on non-relativistic neutrino species is  in  $D_{\nu} (k,z)$, given by 
\begin{equation}
D_{\nu_i} (k,z) \propto (1-f_{\nu_i}) D(z)^{1-p_i}
\end{equation} 
where $k \gg k_{{\rm fs},i} (z)$ and $p_i = (5- \sqrt{25 - 24
  f_{\nu_i}})/4$. 
The standard linear growth function $D(z)$ fitting formula is taken
from \cite{HuEise}.  
$\Delta$ directly changes the growth on scales $k_{{\rm fs},m}\!<\!k\!<k_{{\rm fs},\Sigma}$, i.e., 
$\sqrt{(1-\Delta)/3}< k/(k_0 \sqrt{\Sigma})<1$, which leads to changes in the power spectrum 
that can be aproximated at linear order by 
\begin{equation}
\frac{P(k;z)_{\nu}-P(k;z)}{P(k;z)} \propto - f_{\nu,\Sigma} (1-\Delta)/3  \,.
\end{equation}

 \begin{center}
\begin{figure}[]
\hspace*{-0.1cm}
\includegraphics[width=9cm]{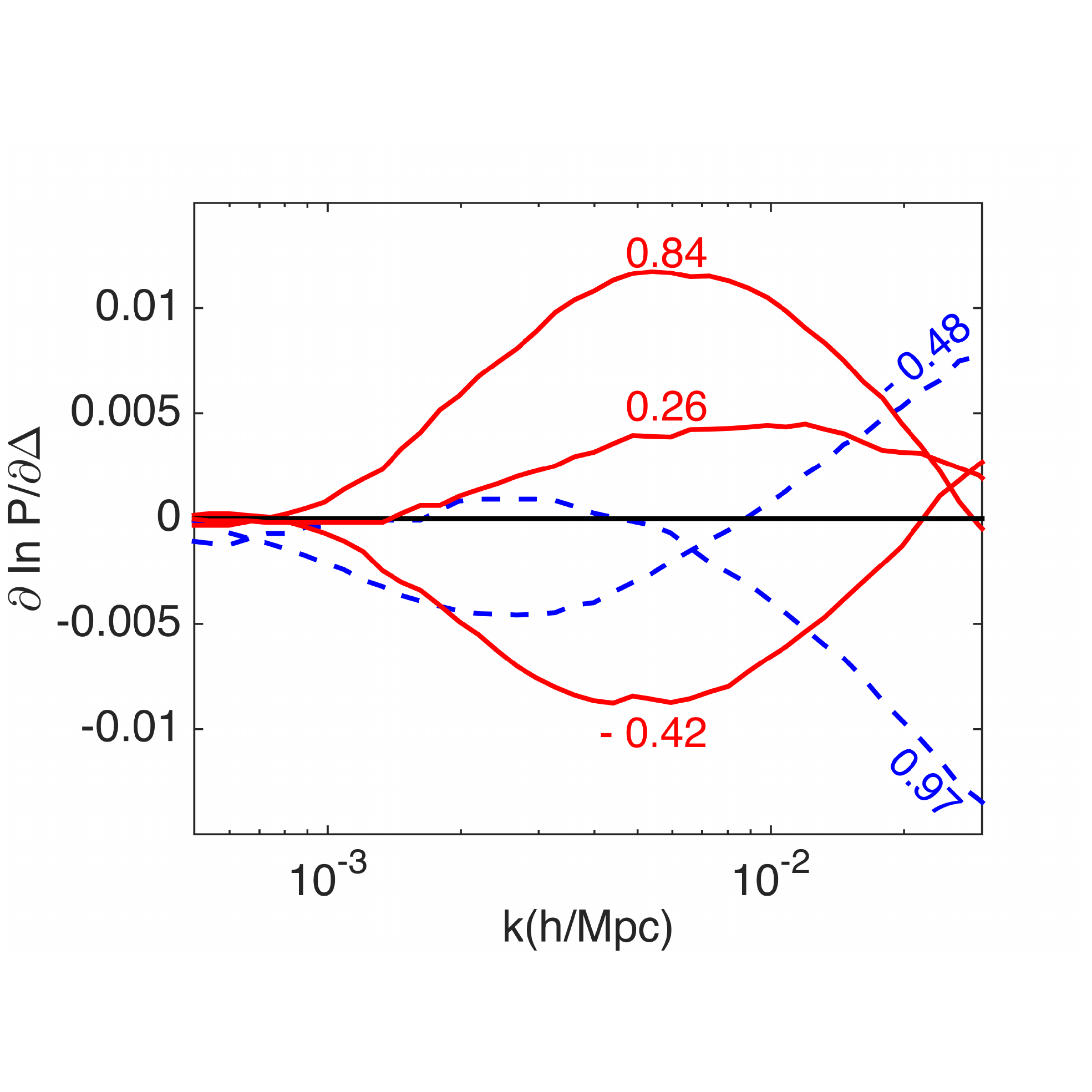}
\vspace*{-1.5cm}
\caption{Dependence of $P(k)$ on the parameter $\Delta$ at $z=0$, for
  fixed $\Sigma$, 0.1 eV (red full) and 0.06 eV (blue dashed), and several values of $\Delta$. The dependence is
  expressed as fractional variation in $P(k)$ for a unit variation in
  $\Delta$. For $\Sigma=$0.1 eV, normal
  (inverted) hierarchy from oscillation data correspond to $\Delta \sim 0.3~(-0.5)$, as shown in Fig.1.  The changes visible at $k>10^{-2}$h/Mpc for $\Sigma=0.06$ eV are due to $\Delta$ changing the matter-radiation equality and thus the shape parameter, being all other cosmological parameters kept fixed. }
\label{fig:dpddelta}
\end{figure}
\end{center}

Firstly, $\Sigma$ and $\Delta$ modify the range of scales where the impact of $\Delta$ is important. 
In particular, smaller total mass of neutrino leads to a narrower range of larger scales where $\Delta$ influences 
the power spectrum. Secondly, while $\Delta = 0$  leads to lack of a signature, positive (negative) $\Delta$ increases (decreases) 
linearly the amount of power at the scales of influence. 

This description is, however, incomplete: the transition between the
different  
regimes is done sharply in $k$ while in reality the change is very
smooth. 
In addition, the individual masses change the details of the
matter-radiation 
transition which (keeping all other parameters fixed) adds an
additional effect at scales $k>k_{{\rm fs},\Sigma}$.  In what follows we will therefore use the full numerical evaluation \cite{camb}.

\begin{center}
\begin{figure}[]
\hspace*{-1.cm}
\includegraphics[width=8cm]{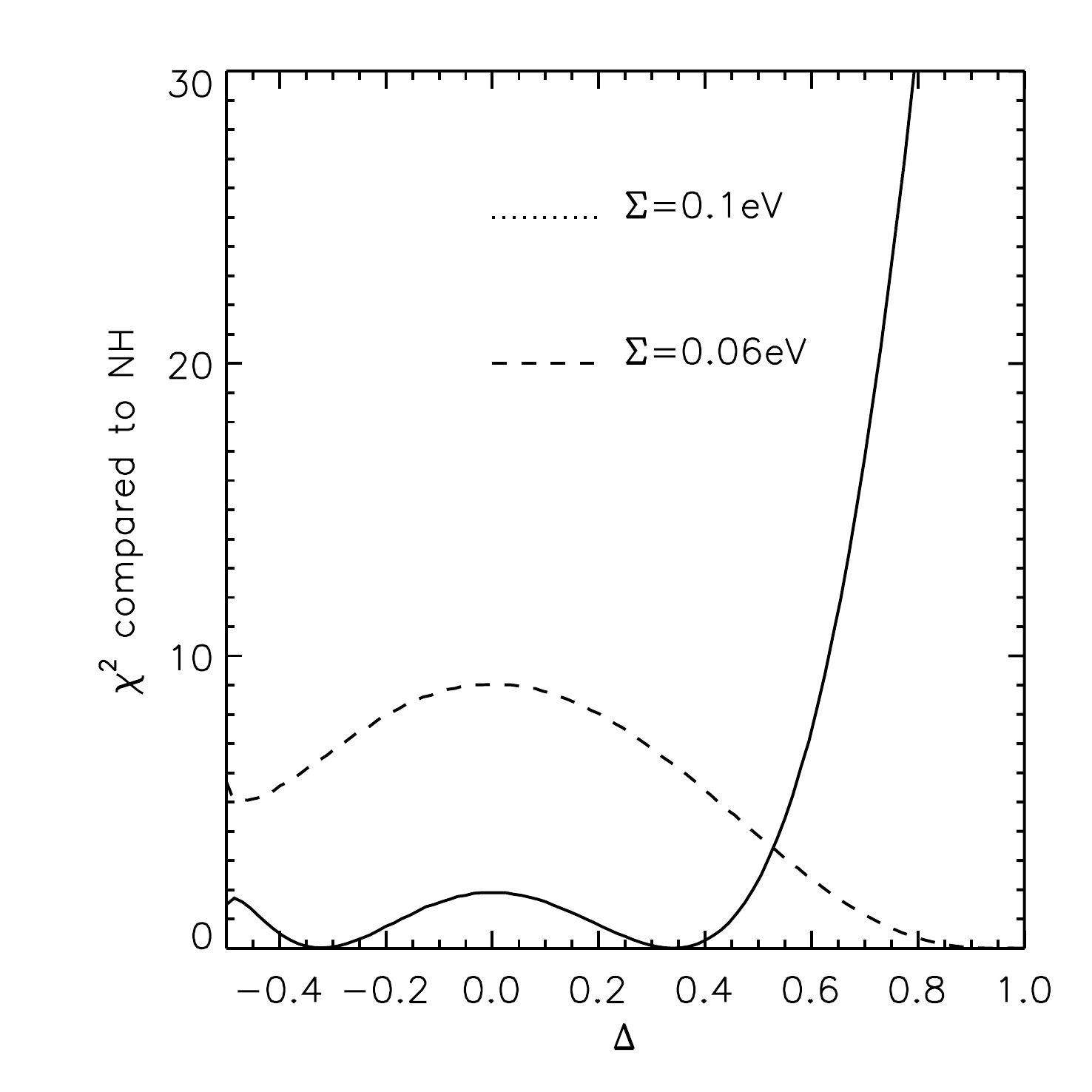}
\caption{Statistical power of future surveys, described in \cite{paper1}, to determine $\Delta$ for fixed values of $\Sigma$. We assume normal hierarchy fiducial model, with $\Delta$ from neutrinos oscillation data,
Many $\Delta$ values potentially inferred from cosmological data can be statistically distinguished from the fiducial model.}
\label{fig:chisq}
\end{figure}
\end{center}

Throughout this paper we assume a fiducial model given by basic
parameters of the standard LCDM 
cosmology and the fiducial values for $\Sigma$ and
$\Delta$ 
are then further specified in each case. We have used the publicly available 
{\tt CAMB} code \cite{camb} to numerically compute the matter power spectrum.
In Fig.~\ref{fig:dpddelta} we show the dependence of $P(k)$ on the
parameter $\Delta$ at $z=0$ for fixed $\Sigma$ and fixed
cosmological parameters. The dependence is shown as the  
 fractional change of the matter power spectrum for a  unit change of
 the parameter $\Delta$. 
  In order  to compute reliably the above derivatives, 
{\tt CAMB} needs to be run at the highest precision settings, 
with fine $k$ sampling and taking care that interpolations 
procedures in-built in the code do not  introduce a spurious signal.  Note that the effect appears on large, linear scales. Moreover, the location, trend and amplitude of the maximum deviation of $\partial \ln P/\partial \Delta$ depend on $\Sigma$:  oscillation experiments constraints on mass splitting  together with a value of  $\Sigma$ completely specify the expected signal.  In Fig. \ref{fig:chisq} we estimate the expected signal to noise  in terms of difference in $\chi^2=-2\ln L$ as a function of $\Delta$   fixing all other cosmological parameters, for a survey with the same characteristics as considered in \cite{paper1}. In this calculation we considered only scales $k<k_{\rm max}^*$ where $k_{\rm max}^*$ represent the scale  above which $\partial \ln P /\partial \Delta$ becomes constant (and non-zero): by fixing all cosmological parameters, $\Delta$ alters  matter  radiation equality  and so the  $\Gamma$  shape parameter  and the $P(k)$ normalisation at scales  below equality, these changes would be canceled through cosmological degeneracies with other parameters (e.g.,  
 for  $\Sigma=0.1$eV  $k_{\rm max}^*=2.5\times 10^{-2}$ h/Mpc).
 
 In other words, if for example  the small scale power suppression signal indicates a $\Sigma=0.1$ ($0.06$)eV, the large scale measurement should yield  a value consistent with $|\Delta|\sim 0.3$ ($\Delta\sim 1$). Excluding other values such as   $\Delta=1$ ($\Delta=0$ or $\Delta<0$), which, as Fig.~\ref{fig:chisq}  shows, future surveys have the statistical power for achieve,   would offer a powerful consistency check on the $\Sigma$ determination. Conversely, excluding the expected value for $\Delta$ at high confidence would cast doubts on  the interpretation of the data in terms of neutrino masses and standard neutrino properties \cite{inprep}.

To summarize, we have shown the existence of  a well defined footprint of neutrinos in the matter power spectrum 
if the total mass of neutrinos is measured in the upcoming cosmological data.  This footprint is localised on large, linear scales,  where systematic effects that plague the $\Sigma$-sensitive small mildly non-linear scales (non-linearities, non-linear bias, baryonic and astrophysical effects, shot noise etc.) are not present: the two signatures are not independent (in the standard model they are both made by neutrinos) but, because of the separation of scales, statistically uncorrelated. They could be partially correlated through cosmological parameter degeneracies but this is left for future work. The systematic effects at play on these large  scales  (i.e., modelling of the  survey window function, selection function, photometric redshifts  errors, if applicable, general relativistic effects, primordial-non-gaussianity) are completely independent form those acting at small scales and more benign, moreover they show a  dependence on k-scale  fundamentally different from the signal. One may wonder on the implications of this result. 
For example,  not finding the power suppression should erase the footprint we discussed here, but will lead to modifications of either the cosmological model or the neutrino properties. Understanding the new physics beyond the standard (cosmology or neutrino) model would be  within reach with improvements in neutrino beta and double-beta decay experiments and further cosmological surveys.
Therefore a detection of  this large-scale signature or  a null result will have profound implications in neutrino physics and cosmology.  A detection will be 
 a ``smoking gun"  for  verifying power suppression of small scales by neutrinos.  As such it  will be an indirect discovery of relic neutrinos and lead to the stronger bound on properties of neutrino like neutrino invisible decays, limiting very small neutrino couplings to scalars g $ \lesssim 4\cdot10^{-14}$ \cite{Serpico:2007pt}.

\acknowledgments{Funding for this work was partially provided by the Spanish MINECO under projects FPA2014-57816-P, AYA2014-58747-P and MDM-2014- 0369 of ICCUB 
(Unidad de Excelencia Maria de Maeztu), by Generalitat Valencia under Prometeo Grant II/2014/050 and by PITN- GA-2011-289442-INVISIBLES.

\end{document}